%% file: parallel-DQC.tex
\documentclass[aps,prx,superscriptaddress,twocolumn]{revtex4-2}
\usepackage[utf8]{inputenc}
\usepackage{bm}
\usepackage{bbm}
\usepackage{bbold}
\usepackage{amsfonts}
\usepackage{amsmath}
\usepackage{amssymb}
\usepackage{graphicx}
\usepackage{mathtools}
\usepackage{upgreek}
\usepackage{xfrac}
\usepackage{soul}
\usepackage{xcolor}

\usepackage[colorlinks=true,citecolor=blue]{hyperref}
\hypersetup{colorlinks=true,citecolor=blue,linkcolor=blue,urlcolor=blue}
\usepackage{url}

\usepackage{txfonts}
\DeclareSymbolFont{matha}{OML}{txmi}{m}{it}
\DeclareMathSymbol{\varv}{\mathord}{matha}{118}

\makeatletter
\newsavebox{\@brx}
\newcommand{\llangle}[1][]{\savebox{\@brx}{\(\m@th{#1\langle}\)}%
  \mathopen{\copy\@brx\kern-0.5\wd\@brx\usebox{\@brx}}}
\newcommand{\rrangle}[1][]{\savebox{\@brx}{\(\m@th{#1\rangle}\)}%
  \mathclose{\copy\@brx\kern-0.5\wd\@brx\usebox{\@brx}}}
\makeatother

\begin{document}

\title{Physics-Informed Quantum Machine Learning: Solving nonlinear differential equations in latent spaces without costly grid evaluations}

\author{Annie E. Paine}
\affiliation{Department of Physics and Astronomy, University of Exeter, Stocker Road, Exeter EX4 4QL, United Kingdom}
\affiliation{PASQAL, 7 Rue Léonard de Vinci, 91300 Massy, France}

\author{Vincent E. Elfving}
\affiliation{PASQAL, 7 Rue Léonard de Vinci, 91300 Massy, France}

\author{Oleksandr Kyriienko}
\affiliation{Department of Physics and Astronomy, University of Exeter, Stocker Road, Exeter EX4 4QL, United Kingdom}
\affiliation{PASQAL, 7 Rue Léonard de Vinci, 91300 Massy, France}

\date{\today}

\begin{abstract}
We propose a physics-informed quantum algorithm to solve nonlinear and multidimensional differential equations (DEs) in a quantum latent space. We suggest a strategy for building quantum models as state overlaps, where exponentially large sets of independent basis functions are used for implicitly representing solutions. By measuring the overlaps between states which are representations of DE terms, we construct a loss that does not require independent sequential function evaluations on grid points. In this sense, the solver evaluates the loss in an intrinsically parallel way, utilizing a global type of the model. When the loss is trained variationally, our approach can be related to the differentiable quantum circuit protocol, which does not scale with the training grid size. Specifically, using the proposed model definition and feature map encoding, we represent function- and derivative-based terms of a differential equation as corresponding quantum states. Importantly, we propose an efficient way for encoding nonlinearity, for some bases requiring only an additive linear increase of the system size $\mathcal{O}(N + p)$ in the degree of nonlinearity $p$. By utilizing basis mapping, we show how the proposed model can be evaluated explicitly. This allows to implement arbitrary functions of independent variables, treat problems with various initial and boundary conditions, and include data and regularization terms in the physics-informed machine learning setting. On the technical side, we present toolboxes for exponential Chebyshev and Fourier basis sets, developing tools for automatic differentiation and multiplication, implementing nonlinearity, and describing multivariate extensions. The approach is compatible with, and tested on, a range of problems including linear, nonlinear and multidimensional differential equations.  
\end{abstract}

\maketitle

\section{Introduction}

Fast, flexible and efficient solvers of differential equations (DEs) underpin the area of scientific computing \cite{HeathSciComp}. Applied to computationally hard problems in physics \cite{braun1983differential}, fluid dynamics \cite{anderson1995computational}, geoscience, chemistry \cite{jensen2013introduction}, and finance \cite{oksendal2010stochastic}, differential equation solvers consume a significant portion of the world's high-performance computing resources \cite{ExaScl,ExaDG2020}. The quest for developing and improving DE solvers remains open. Current classical approaches use variants of finite element methods (FEM) \cite{leveque2007finite} for formulating problems as a linear algebra task and employing matrix inversion. Alternatively, spectral element methods (SEM) are used \cite{boyd2013chebyshev,hussaini1983spectral}. While FEM and SEM solvers are robust and suit many cases \cite{Knio2006}, they are computationally intense for large grids and large basis sets, suffering from the curse of dimensionality \cite{Gear1981}. For instance, considering $L$ points for the FEM grid and $d$ dimensions, in the worst case we need to deal with functions evaluated at $\mathcal{O}(L^d)$ points. Other significant challenges correspond to problems that involve strong nonlinearity, turbulence, stiffness \cite{Gear1981}, stochasticity and strong correlations \cite{Dolgov2020}, optimal control \cite{speyer2010optimal}, and data-driven tasks. 

A qualitatively distinct family of protocols has emerged recently where DE problems are reformulated as a machine learning (ML) task. The corresponding field is referred to as scientific machine learning (SciML) \cite{SciML,rackauckas2019diffeqflux,rackauckas2020universal}. SciML approaches are often based on physics-informed neural networks (PINNs) \cite{Raissi2019PINNs,Cai2021pinn-review} --- deep neural networks that include differential constraints introduced with automatic differentiation. PINN-based solvers have been applied to many use-cases \cite{Raissi2020Science,Brunton2020,DANDEKAR2020,Cuomo2022sciML,Sel2023}. Although still remaining relatively niche \cite{Markidis2021pinns}, their flexibility allows them to compete with FEM/SEM solvers in cases involving data-based constraints \cite{Cuomo2022sciML}, optimization loops \cite{AbuKhalaf2005,Han2018sdes}, and can excel when dealing with stiff and multidimensional problems \cite{Yang2021bpinns,rackauckas2020universal,Kim2021stiff,Sharma2023}. On one end, the training grid for PINN models that generalize well can use $L_c$ training points where $L_c < L$, with savings relevant at large $d$ \cite{Han2018sdes}. However, PINN-based DE solvers require a costly training process that leads to large overheads as compared to conventional methods such as FEM, often needing $>10,000$ epochs with many function and gradient evaluations. While the required function evaluations on the grid may be parallelized on GPUs \cite{rackauckas2020universal}, in practice the cost is often not considered worth it for many industrial applications, despite the potential of the ML setting and what the flexibility can offer in terms of utility in use cases. Reducing the grid evaluation cost for each epoch, and removing $L_c$ scaling, is crucial for making PINNs industrially practical.

Quantum computing \cite{nielsen2010quantum} offers a promise to speed up scientific computing and machine learning. It has unique advantages in generalization \cite{Caro2022,Huang2022} and sampling \cite{Liu2018,Zoufal2019,Coyle2020}, and for random sampling tasks has recently demonstrated quantum supremacy \cite{Arute2019,morvan2023phase}. In regards to differential equations \cite{Balducci2022rev}, quantum computing has often been considered as an accelerator of linear algebra subroutines due to fast matrix inversion, assuming oracle-based input and full statevector readout \cite{Biamonte2017}. In this case, the problem is typically discretized on a grid of $2^N$ points (where $N$ is the number of qubits). Thus, large-$L$ grids can be compressed into $\mathcal{O}(\log L)$ qubits, potentially avoiding the curse of dimensionality. The seminal example of inversion-based linear equation solver corresponds to the HHL protocol \cite{harrow2009quantum} that implements exponentially-improved eigenvalue inversion for discretized operators and amplitude encoded states \cite{Montanaro2016,Costa2019,Linden2022quantumvsclassical}. Other improved algorithms include inversion based on linear combination of unitaries (LCU) \cite{childs2017quantum}, quantum signal processing \cite{Lin2020optimalpolynomial,Wang2022energylandscapeof}, adiabatic inversion \cite{subacsi2019quantum,Costa2022,jennings2023efficient} etc. Spectral-like methods were also considered \cite{Childs2020}. For problems involving time evolution approaches based on quantum lattice Boltzmann solvers \cite{schalkers2022efficient,schalkers2023importance} and Schrodingerisation \cite{jin2022quantum,jin2022quantumEXT,jin2023quantum} were put forward. Moreover, several solvers for stochastic differential equations were considered \cite{Kubo2021,Alghassi2022variationalquantum}. We note that typically all aforementioned approaches based on amplitude-encoded functions tackle linear problems, and use discretized derivative representation. Treating nonlinear differential equations on quantum computers remains a challenge. To date they were approached via linearization techniques \cite{JPLiu2021,Xue2022nonlinearquadratic,Krovi2023improvedquantum} (Carleman and others), or significantly extending register size \cite{leyton2008quantum,Dodin2021,lloyd2020quantum}. These approaches however are limited to weakly nonlinear problems or require the number of qubit registers to increase directly proportional to the degree of nonlinearity, respectively. 

A parallel stream of research in differential equations has emerged based on hybrid quantum-classical methods and quantum machine learning (QML) techniques. In this case, the goal is often to address genuinely nonlinear problems and offer potential near-term solutions. These include works that implement quantum subroutines for solving fluid dynamics \cite{Gaitan2020,Oz2023}, tensor network-inspired variational protocols with quantum nonlinear processing units \cite{lubasch2020variational,GarciaRipoll2021quantuminspired,Garcia-Molina2022,jaksch2022variational}, and annealing-based approaches \cite{Zanger2021quantumalgorithms,Dukalski2022,Criado2023}. From the SciML perspective, an approach based on introducing nonlinear quantum models and derivative quantum circuits was initially proposed in Ref.~\cite{kyriienko2021solving}. Often referred to as a DQC-based solver, it resembles the PINN approach in classical ML, and can readily address nonlinear differential equations. Other works advanced along these lines, including the model discovery \cite{heim2021quantum}, continuous variable-based workflow \cite{Markidis2022qpinns,Knudsen2020}, stochastic differential equation solvers \cite{Paine2021,kyriienko2022protocols,kasture2022protocols}, solving integro-differential equations and treating integral transforms over the solution space \cite{Kumar2022}, optimizing over differentiable quantum model solutions \cite{Varsamopoulos2022}, and harmonic quantum neural networks for multisimulation \cite{ghosh2022harmonic}. Also, kernel-based quantum DE solvers were developed \cite{paine2023quantum}. The positive sides of QML-based DE solvers are the ability to address nonlinear problems \cite{kyriienko2021solving}, easy readout without tomography, and implementations favoring near-term operation. Being heuristic much like PINNs, the same downside is a non-convex learning process with rugged optimization landscapes \cite{you2021exponentially,Anschuetz2022}, and barren plateaus for non-tailored variational quantum circuits \cite{mcclean2018barren}. A major cost is the loss evaluation, which for large training grids $L_c$ may prevent scaling to large problem sizes. 

In this work, we propose a quantum algorithm to solve differential equations within physics-informed SciML paradigm, where the costly grid evaluation is avoided. This is achieved by expressing a trainable model as an overlap between a state that encodes independent variables and an adjustable quantum state (or mixture of states). We show how to prepare the model via feature map circuits, differentiate them automatically by applying operators that are basis-aware, and introduce a latent space multiplication function. This allows implementing nonlinearity efficiently, with only a modest additive system size increase that scales linearly with the degree of nonlinearity. At the practical level, two exponentially-large basis sets are considered and implemented, corresponding to Fourier \cite{kyriienko2022protocols} and Chebyshev \cite{williams2023quantum} encodings. Crucially, with the chosen model definition, we recast the implementation of differential constraints into loss function minimization that compares distances between quantum states (latent space representations) that stand for functions and derivatives, and does not depend on $L_c$. We also show how mappings between bases allow implementing initial and boundary terms, as well as data-based constraints. The loss can be approached within a hybrid quantum-classical workflow as a non-convex optimization problem, or with quantum inversion-based methods. Using the proposed approach in the variational form, we tested it for various problems (linear, multidimensional, nonlinear) and confirmed that high-quality solutions can be recovered.

The paper is structured as follows. First, we introduce the protocol \cite{patent} for latent space-based differential equation solving in qualitative terms, and present technical descriptions for general choice of encoding. We then explore two specific encoding choices --- Chebyshev and Fourier. Finally, results from the simulation of the protocol are described.
\begin{figure}
    \begin{center}
    \includegraphics[width=1.0\linewidth]{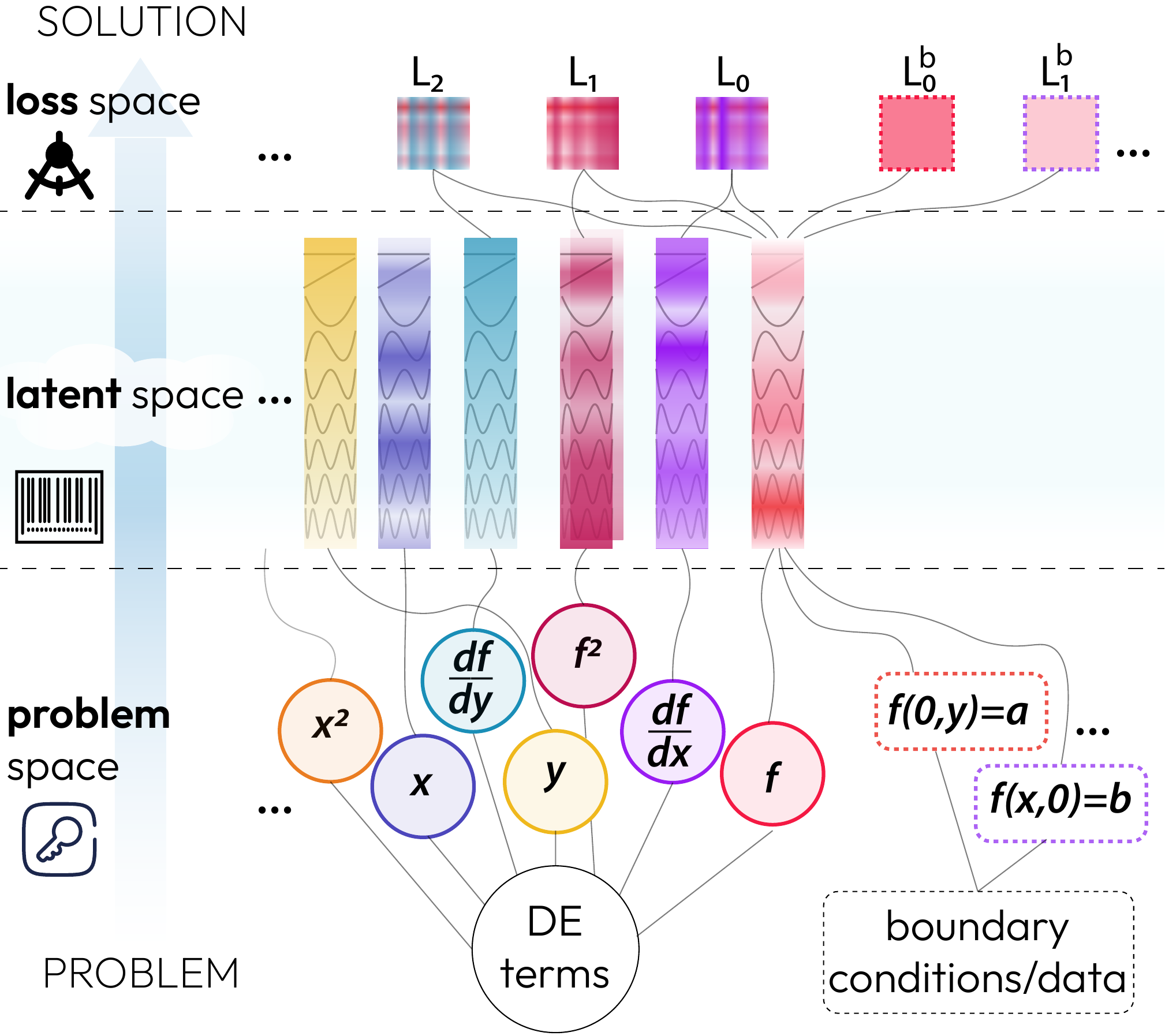}
    \end{center}
    \caption{A conceptual visualization of the proposed physics-informed approach to solving nonlinear differential equations (bottom-up description). We start from the problem space, where differential equation (DE) terms, their relations, and the boundary terms are provided. Next, these terms are elevated to the latent space via feature map that prepares corresponding quantum states, with amplitudes associated to components of independent basis functions. These states are then compared via pairwise overlaps, forming corresponding distances in the loss space, and boundary terms are evaluated explicitly via basis mappings. The total loss can be minimized to learn a model that provides a solution to the problem.}
    \label{fig:concept}
\end{figure}


\section{The protocol}

\subsection{Generic overview}

Our aim is to solve differential equations by encoding their solutions into quantum models and, broadly speaking, utilizing a quantum parallelism \cite{nielsen2010quantum}. The protocol employs physics-informed constraints for learning, and efficiently evaluates loss terms on an entire computational grid (specifically, for the full set of independent basis functions). While the full story is rather involved, here we shall first try to describe it in an intuitive way.

Imagine that you need to solve a differential equation for some function $f(x)$ of a scalar variable $x$. The equation involves derivatives ($df/dx$, $d^2f/dx^2$ etc), and it also may include the function itself, its square, plus other functions that depend on $x$. This can be readily extended to multidimensional problems that include other independent variables ($y$ etc). Additionally, the problem of solving differential equations assumes specifying initial or boundary terms. Say, we want to solve $df/dx-f(x)=0$. How can we check if the derivative indeed matches the function itself? We posit that this can be addressed as a global evaluation of compatibility of the two DE terms in a \emph{latent space}.

Let us imagine a collection DE terms as the problem space (Fig.~\ref{fig:concept}, ground layer). Our next step is to associate each term with a quantum state living in an $N$-qubit Hilbert space that spans $2^N$ states. We refer to this as a the latent space, where DE terms are elevated to (Fig.~\ref{fig:concept}, middle layer). For this, we develop a recipe (mapping $f \rightarrow |f\rangle$, $df/dx \rightarrow |f'\rangle$, $x \rightarrow |x\rangle$ etc) such that each term in the problem space has its own quantum representation. Here, we make sure that different DE terms can be projected in the same basis (for instance, shown as a set of Chebyshev polynomials in Fig.~\ref{fig:concept}), and the corresponding functions can be read out as overlaps with states representing variables (i.e. $|x\rangle$). Next, we need to check if the constraints hold. Returning back to the simple example above, we compare the terms $df/dx$ and $f(x)$ in the same basis by measuring the effective distance between latent space representations, corresponding to the overlap of quantum states $|f'\rangle$ and $|f\rangle$. This is repeated for all terms in the problem, and our goal is learn the solution that satisfies constraints and initial values (or some data). Here, we can write a total loss as a sum of individual contributions $L_i$ (Fig.~\ref{fig:concept}, top layer) and learn the solution variationally for an adjustable model $f_\theta(x)$, similarly to classical physics-informed neural network approach and quantum approaches for derivative quantum circuits \cite{kyriienko2021solving}. Alternatively, we can use the evaluated constraints to form a system of equations and solve it using quantum matrix inversion techniques.

The protocol can be also imagined as a hashing procedure \cite{knuth1998art}. Data is supplied in the form of DE terms representing ``keys''. Applied quantum embedding transforms them into latent space representation, which can be seen as effective ``barcodes''. Finally, distances between barcodes are measured, where hashing produces a set of scalars that are combined into the total loss.

The next question is: how can we formalize this procedure?


\subsection{Technical overview}
We proceed with a formal definition of the problem. The goal of the protocol is to solve a system of $D$ differential equations,
\begin{align}
\label{eq:DE-problem}
    \mathrm{DE}_j(\bm{x},\bm{f},d\bm{f}/d\bm{x}, d^2\bm{f}/d\bm{x}^2, ...) = 0 ~~ \mathrm{for}~j=1:D,
\end{align}
where $\mathrm{DE}_j(\bm{v})$ denotes a sum of terms $\bm{v}$ in $j$-th differential equation that equates to zero. Additionally, the initial or boundary conditions have to be specified and satisfied. We want to obtain a solution of Eq.~\eqref{eq:DE-problem} as functions $\bm{f}(\bm{x})$ of variables $\bm{x}$. For this, we use a quantum model $\bm{f}_{\bm{\theta}}(\bm{x})$ parametrized by a vector of weights $\bm{\theta}$, and reformulate the problem as learning an optimal model $\bm{f}_{\theta^*}(\bm{x})$ to represent $\bm{f}(\bm{x})$. Specific to our protocol, we set up the model in the form that can be evaluated at any value of variable $\bm{x}$, while also admitting the latent space representation such that the learning process does not require explicit evaluation of the quantum model at the grid of training points. To keep the discussion simple, initially we consider the case with $D=1$, omitting the sub-indices ($\mathrm{DE}_{j=1} \equiv \mathrm{DE}$), keeping first-order derivatives only, and a single dimension for both dependent and independent variables, i.e. $\mathrm{DE}(x,f,df/dx) = 0$. The details of generalization are described in the last subsection of the protocol section. 

The first step is to note that the $\mathrm{DE}$ can be written as
\begin{align}
    \label{eq:DE_sum}
    \mathrm{DE}(x, f,df/dx) = \sum_{k=1}^{T} \mathrm{DE}^{[k]}(x,f,df/dx) = 0,
\end{align}
where $\{ \mathrm{DE}^{[k]}(x,f,df/dx) \}_k$ denote the separate terms of the differential equation formed of products of $f$, and $df/dx$ as well as arbitrary functions of $x$. We instantiate a quantum model in the general form
\begin{align}
    \label{eq:f-model}
    f_\theta(x) = \langle x | f_\theta \rrangle,
\end{align}
represented by the overlap between a quantum state $|x\rangle$ and a \emph{mixture} of quantum states $|f_\theta \rrangle$. We define the mixture as a sum of quantum states weighted with individual coefficients (as for instance used in quantum filter diagonalization \cite{parrish2019quantum,Bespalova2021prx,Kyriienko2020npj}), and label it using the double-ket notation, $|\circ \rrangle$. We stress that single-state models are naturally included into the mixture-based models, $| f_\theta \rangle \in | f_\theta \rrangle$, and can be readily used. We also introduce the state $|x\rangle$ that encodes the variable $x$, in particular using the computational basis. Importantly, we can generate $|x\rangle = \hat{\mathcal{U}}(x)|{\o}\rangle$ by applying an $x$-dependent unitary operator to the computational zero state $|{\o}\rangle$ \cite{kyriienko2022protocols,williams2023quantum}. Here $\hat{\mathcal{U}}(x)$ represents a feature map \cite{mitarai2018quantum,Schuld2019feature}. Within the proposed quantum model definition, we posit that each term $\mathrm{DE}^{[k]}$ can be expressed as
\begin{align}
\label{eq:DE_term}
    \mathrm{DE}^{[k]}(x,f_\theta,df_\theta/dx) &= \langle \tilde{x} | \mathrm{DE}_\theta^k \rrangle , \\
    | \mathrm{DE}_\theta^k \rrangle &= \sum_{\ell=1}^{t_k} \alpha^k_{\ell} |\Delta_\theta^{k, \ell}\rangle,
    \label{eq:term_exp}
\end{align}
where $|\tilde{x}\rangle$ is a quantum state corresponding to the latent space representation of the variable. We note that $|\tilde{x}\rangle$ is the same for all terms when forming the model. However, it may differ from $|x\rangle$ and imply an alternative basis. This will be specifically required when dealing with product terms later. In Eq.~\eqref{eq:term_exp} we explicitly show that each term of the differential equation $\mathrm{DE}^{[k]}$ can be based on a sum of $t_k$ parametrized quantum states $\{ |\Delta_\theta^{k, \ell}\rangle \}_{\ell=1}^{t_k}$, weighted by coefficients $\{ \alpha^k_{\ell} \}_{\ell=1}^{t_k}$. Here, we keep the mixture states $| \mathrm{DE}_\theta^k \rrangle$ in the general form, and provide the details for their construction in the following section. 
Substituting the expression in Eq.~\eqref{eq:DE_term} into Eq.~\eqref{eq:DE_sum} we get
\begin{align}
    \label{eq:de_sep}
    \mathrm{DE}(x,f_\theta,df_\theta/dx) = \langle \tilde{x} | \sum_{k=1}^{T} | \mathrm{DE}_\theta^k \rrangle = 0, 
\end{align}
for all $x$. Essentially, Eq.~\eqref{eq:de_sep} represents a check of $T$ differential constraints for the model \eqref{eq:f-model} based on quantum state overlaps with the same state $|\tilde{x}\rangle$. We additionally restrict the state $|\tilde{x}\rangle$ such that its amplitudes form an independent basis set (discussed in detail in the next subsection). 
With this restriction the differential equation is satisfied if and only if $\sum_{k=1}^{T} | \mathrm{DE}_\theta^k \rrangle = \bm{0}$, where the used notation implies that every element of the differential constraint vector has zero value. If the restriction on $|\tilde{x}\rangle$ is not in place then $\sum_{k=1}^{T} | \mathrm{DE}_\theta^k \rrangle = \bm{0}$ is sufficient but not a necessary condition. In this case solutions can be missed when training for $\sum_{k=1}^{T} | \mathrm{DE}_\theta^k \rrangle = \bm{0}$. 
Therefore, by constructing $|\mathrm{DE}_\theta^k \rrangle$ for each differential equation term with adjustable parameters the solution can be learnt by solving $\sum_{k=1}^{T} | \mathrm{DE}_\theta^k \rrangle = \bm{0}$ in several ways.

\textit{Loss-based variational optimization.---}For solving this problem we first consider a variational approach based on minimizing the loss function
\begin{align}
\label{eq:loss_DE}
\mathcal{L}_{\mathrm{DE}}(\theta) &= \sum_{k=1}^{T} \llangle \mathrm{DE}_\theta^k| \sum_{m=1}^{T} | \mathrm{DE}_\theta^m \rrangle = \\
&= \sum_{k=1}^{T} \sum_{m=1}^{T} \sum_{\ell=1}^{t_k} \sum_{n=1}^{t_m} \alpha^k_\ell \alpha^m_n\langle \Delta^{k,\ell}_\theta| \Delta^{m,n}_\theta \rangle.
\end{align}
To evaluate the loss function $\mathcal{L}_{\mathrm{DE}}(\theta)$ the overlaps between prepared quantum states are measured and then post-processed classically to account for weights. Similarly, we need to evaluate derivative-based contributions. Once states are prepared, the overlaps are estimated using known methods, for instance the Hadamard test \cite{Mitarai2019Htest} and other techniques typically used in quantum kernel methods \cite{Schuld2019feature,paine2023quantum} or ground state estimation \cite{Kyriienko2020npj,Bespalova2021prx}. We provide a brief summary and circuits for evaluating overlaps in Appendix \ref{app:A}.

Next, we need to introduce the boundary or initial conditions for solving differential equations. The significance of this step shall not be underestimated, as differential constraints can be satisfied in many ways, and it is the initial value terms that pin the solution. Similarly, one can imagine problems where a set of data is supplied to regularize $f_{\theta}(x)$, and we need to include this information when learning the solution. We resolve the question of introducing initial value terms by utilizing the access to our model via quantum feature map circuits. Namely, the function can be evaluated at some initial point $x_0$ as $f_\theta(x_0) = \langle {\o}| \hat{\mathcal{U}}(x_0)|f_\theta \rrangle$, rather than globally. For instance, given the initial value condition $f(x_0) = f_0$, we can introduce a corresponding loss term as a distance measure between $f_\theta(x_0)$ and $f_0$, which can be additionally weighted by a pre-defined constant $\eta$. Using the mean square error (MSE) for defining the distance, this loss term is
\begin{align}
    \mathcal{L}_{\mathrm{init}}(\theta) = \eta \left\{f_\theta(x_0)-f_0 \right\}^2,
\end{align}
where $\eta$ controls the importance of the initial value contribution. 
Similarly, we can include data dependence for regularizing the solution as 
\begin{align}
    \mathcal{L}_{\mathrm{data}}(\theta) = \zeta \sum_{x_i \in \mathcal{X}} \left\{f_\theta(x_i) - f_i \right\}^2,
\end{align}
where $\mathcal{X} = \{x_i\}_i$ are grid points in the supplied data set, with corresponding function values $\{f_i\}_i$, and $\zeta$ being a weighting parameter.

The total loss is then
\begin{align}
    \mathcal{L}(\theta) = \mathcal{L}_{\mathrm{DE}}(\theta) + \mathcal{L}_{\mathrm{init}}(\theta) + \mathcal{L}_{\mathrm{data}}(\theta),
\end{align}
which is to be minimized either by non-convex optimization methods, finding $\theta^* = \mathrm{argmin}_{\theta} \mathcal{L}(\theta)$, or applying other iterative methods to recover the optimal mixture state $|f_{\theta^*}\rrangle$. When choosing the variational optimization, it is convenient to use a gradient descent and adjust angles $\theta$ based on $\nabla_{\theta} \mathcal{L}(\theta)$. This can be approached by various circuit differentiation techniques, including the parameter shift rule \cite{Schuld2019parametershift,mitarai2018quantum} and generalizations for wider range of circuits \cite{kyriienko2021generalized,Izmaylov2021,Wierichs2022generalparameter}, or LCU-based derivatives that are specifically fitting the differentiation of kernels \cite{paine2023quantum,Mitarai2019Htest}.

Second, we propose a matrix inversion-based workflow that avoids variational approaches and exploits the same problem encoding in the latent space. This is discussed in Appendix \ref{app:B} where we outline the procedure.

Finally, the overall workflow for the protocol (Fig.~\ref{fig:concept}) is summarized below.
\begin{enumerate}
 \item Choose DE with initial value to solve, thus specifying the problem. Split DE into separate product terms.
 \item Choose hyperparameters and set-up workflow components such as $x$ encoding, mixture state for building the model, optimization procedure etc.
 \item Run circuits to prepare relevant state(s) associated to each term (thus instantiating the model in the quantum latent space).
 \item Compute overlaps between terms (therefore project results back to scalars forming individual loss contributions).
 \item Optimize the total loss until differential and boundary constraints are satisfied.
\end{enumerate}

\subsection{Function representation}

A vital component of the proposed algorithm is being able to prepare $| \mathrm{DE}^k_\theta \rrangle$  for various terms that can exist in a generic differential equation. This preparation largely depends on how we represent the $x$-dependence via feature mapping $|x\rangle = \hat{\mathcal{U}}(x)|{\o}\rangle$. The basis functions of the model depend on how we choose $\hat{\mathcal{U}}(x)$ acting on $N$ qubits. Certain limitations on $\hat{\mathcal{U}}(x)$ are imposed such that states associated with derivatives with respect to $x$, functions of $x$, and products can be prepared. The first restriction on $\hat{\mathcal{U}}(x)|{\o}\rangle$ is such that the generated basis functions are independent. To illustrate this, let us write explicitly the state representing the variable as a vector with amplitudes $\tau_j$ that depend on $x$,
\begin{align}
    \label{eq:basis_fn_x}
    |x\rangle = (\tau_0(x), \tau_1(x), ..., \tau_{2^{N-1}}(x))^\mathrm{T},
\end{align}
and choosing $2^N$ basis functions such that $\{\tau_j\}_j$ are linearly independent, meaning that $\forall ~j ~ \nexists ~ \{c_k\}_k ~ \text{s.t.} ~ \tau_j(x) = \sum_{k \in \bar{j}} c_k \tau_k(x)$. This ensures that Eq.~\eqref{eq:de_sep} is solved if and only if $\sum_{k=1}^{T} | \mathrm{DE}^k \rrangle = \bm{0}$ holds. 

Next, as discussed in the previous subsection, our model is built as the overlap $f_\theta(x) = \langle x| f_\theta \rrangle$. Using the same basis as for encoding $x$, we can express the mixture state as a vector of $2^N$ components,
\begin{align}
    \label{eq:basis_fn_f}
    | f_\theta \rrangle = (f_{\theta,0}, f_{\theta,1}, ..., f_{\theta,2^N-1})^\mathrm{T}.
\end{align}
This leads to the model represented by the sum of the independent basis functions,
\begin{align}
    \label{eq:basis_fn_model}
    f_\theta(x) = \langle x| f_\theta \rrangle = \sum_{j=0}^{2^N-1} f_{\theta,j} \tau_j^*(x),
\end{align}
which allows introducing rules for differentiating the model. 
In this subsection we further discuss how to prepare $|\mathrm{DE}^k_\theta \rrangle$ for the model, its derivatives, other functions of $x$, and products of these functions for general choice of $\hat{\mathcal{U}}(x)|{\o}\rangle$. Later in the text we will present the specific state preparation strategies for two choices of encoding --- orthonormal Fourier and Chebyshev bases \cite{williams2023quantum}.


\textit{Scaled-and-shifted function representation.---}As long as the conditions on the basis are satisfied, we can further extend the function representation introducing additional scaling factors and additive shifts. This is particularly important when we need to increase or limit the expressivity of our models, also impacting the trainability. Here, the expressivity corresponds to the range of functions it is able to represent, and the trainability defines how easy it is to train the model to represent a particular function that can be expressed. To highlight the need for these considerations, we note that in case of choosing the model as $f_\theta(x) = \langle x | f_\theta\rangle$, it is bound in $0\leq |f_\theta(x)| \leq 1$ range, thus being significant limited in expressivity. One option that is readily available corresponds to introducing a scaling factor for the single-state model, which reads
\begin{align}
    \label{eq:f-model_sc}
    f_\theta(x) = \alpha \langle x | f_\theta\rangle,
\end{align}
with $\alpha$ being a parameter (fixed or adjustable). Another option corresponds to adding a constant to the model in the form
\begin{align}
    \label{eq:f-model_sh}
    f^{\mathrm{sh}}_\theta(x) = \alpha \langle x | f_\theta\rangle + \alpha_{\mathrm{sh}} \equiv \alpha \langle x|f_\theta\rangle  + \alpha_{\mathrm{sh}} g_1(x),
\end{align}
where we highlight that adding the constant $\alpha_{\mathrm{sh}}$ is equivalent to introducing the function $g_1(x) \coloneqq 1$ (up to a scaling factor). For this we introduce the state $|\psi_1\rangle$ such that $\langle x |\psi_1\rangle = 1$. We note that the shifted model falls into a category of mixture-based models with $|f_\theta\rrangle = \alpha |f_\theta\rangle + \alpha_{\mathrm{sh}} |\psi_1\rangle$, and even more general operators than scaling and shifting may be applied to produce $|f_\theta\rrangle$. In order to be able to use the shifting strategy, we note that the embedding has to be chosen such that the unity function can be expressed by the basis, and that its latent space representation is either analytically or variationally obtainable.


\textit{Evaluating derivatives.---}To include the differential constraints we need to connect a formal model derivative with the overlap evaluation based on a modified \emph{derivative} (mixture) state. Taking the formal derivative of $f_\theta(x)$ we get
\begin{align}
    f'_\theta(x) &= \frac{d}{dx}\langle x | f_\theta \rrangle= \langle {\o} | \hat{\mathcal{U}}^{\dag \prime}(x) | f_\theta \rrangle,
\end{align}
where we need to evaluate the derivative for the conjugate transposed feature map $\hat{\mathcal{U}}^{\dag \prime}(x)$. We note that for certain $x$-parametrization we can associate a derivative of an operator with the product of two operators, $\hat{\mathcal{U}}'(x) = \check{G} \hat{\mathcal{U}}(x)$, where $\check{G}$ is an $x$-independent operator to be determined. This is the form of embedding we rely upon. As in general the emergent operator (matrix) is not unitary, we use the check mark $\check{\circ}$ to denote this, also marking such operators throughout. In practice, these operations can be readily absorbed into the mixture state definition, or compiled with the linear combination of unitaries (LCU) approach \cite{childs2017quantum}, or other decomposition methods (QR, Householder etc \cite{Malvetti2021quantumcircuits}). For example, taking a feature map of the form $\exp(-i x \hat{H})$ generated by the Hermitian operator $\hat{H}$ we observe that $\check{G} = -i \hat{H}$ is a skew-Hermitian operator. This can be decomposed into Pauli strings and thus a sum of unitaries. Using the developed notation, the derivative of the model can be expressed as
\begin{align}
    f'_\theta(x) &= \langle x | \check{G}^\dag |f_\theta \rrangle \equiv \langle x | f'_\theta \rrangle,
\end{align}
where we have introduced the derivative mixture state $| f'_\theta \rrangle \coloneqq \check{G}^\dag|f_\theta \rrangle$. Crucially, the state $|f'_\theta \rrangle$ now becomes our proxy to the derivative term evaluation, and serves as $|\mathrm{DE}_{\theta}\rrangle$ for introducing $df_{\theta}/dx$ into the total loss.


\textit{Dealing with independent functions.---}Another type of terms that may arise in differential equations is a function of the independent variable $g(x)$, which we need to include into physics-informed constraints. In this case, we have to exploit the same model structure as for the trial function, and postulate that 
\begin{align}
    \label{eq:arb_fn}
    g(x) = \langle x | g \rrangle,
\end{align}
where $| g \rrangle$ is a state that labels the function. Often, it is sufficient to use a single pure state, $| g \rrangle = |g\rangle$, with a scale factor.

There are several ways how the function $g(x)$ can be loaded into the system. First, we can utilize the fact that the basis set supplied by $\hat{\mathcal{U}}(x)$ is based on orthogonal polynomials. If additionally it is orthonormal (as examples considered in the next section), we can use a unitary map $\hat{\mathcal{U}}_\mathrm{T}$ that transforms states between the encoding basis and the computational basis $|\tilde{x}\rangle$ \cite{kyriienko2022protocols,williams2023quantum}. Using the resolution of identity, we can write the function now as the overlap model in the computational basis, $g(x) = \langle x| \hat{\mathcal{U}}_\mathrm{T}^\dagger \hat{\mathcal{U}}_\mathrm{T} | g \rrangle = \langle \tilde{x} | \tilde{g} \rrangle$. The state $| \tilde{g} \rrangle$ can be identified from the vector of evaluations $\{g(x_j) \}_j$ where $\{x_j\}_j$ are the set of points at which $|x\rangle$ maps to $|\tilde{x}\rangle$. From here we can use quantum state preparation algorithms \cite{rattew2022preparing, zhang2022quantum,gonzalezconde2023efficient} to create the required component states of $| \tilde{g} \rrangle$, sans any prefactors $\alpha$ that can be accounted for separately when evaluating the total loss. 

Alternatively, we can also use regression and learn states $|g \rrangle$ for representing functions with another parametrized mixed state $|g_\phi\rrangle$. This can be achieved for instance via quantum circuit learning \cite{mitarai2018quantum}, set up such that $g(x) =\langle x | g_{\phi^*} \rrangle$ for some optimal parameters $\phi^*$. Scaling and shifting factors (naturally included in the mixture state representation) would help to express functions with varying magnitude. Another alternative for the case when the basis functions in Eq.~\eqref{eq:basis_fn_x} are known, is to utilize a classical method for spectral function decomposition \cite{hussaini1983spectral}, and load corresponding coefficients with the quantum state preparation algorithms.


\textit{Multiplying functions.---}Equipped with the knowledge of how to construct various individual functions and derivatives, we proceed to encoding products of functions. Here, the crucial point is that products require: a) loading information about amplitudes of multiplied states; b) keeping track of the basis that must be extended in the nontrivial way. For this, let us consider a minimal example of multiplying two functions $g(x) = \langle x | g\rrangle$ and $h(x) = \langle x | h\rrangle$, based on the corresponding states. We note that the same procedure can be employed to multiply the model $f(x)$ itself, for instance representing powers $f^2(x)$ etc.

As functions $g(x)$ and $h(x)$ both depend on the variable $x$, we can expand them as $g(x) = \sum_j g_j \tau_j(x)$ and $h(x) = \sum_j h_j \tau_j(x)$, where $g_j$, $h_j$ are state amplitudes. Therefore, the product of functions corresponds to  
\begin{align}
\label{eq:prod_fn}
    g(x)h(x) = \sum_{j,k} g_j h_k \tau_j(x) \tau_k(x),
\end{align}
and we observe that in general it also includes products of basis functions. While previously we specified each function in terms of components for the $\mathcal{A} = \{\tau_j(x)\}_j$ bases, now the emerging basis is $\mathcal{A}^{(2)} = \{\tau_j(x) \tau_k(x) \}_{j,k}$ that may not be unique or independent. To fix this, we consider a new basis set $\mathcal{B} = \{\beta_j(x)\}_j$ which is independent, and choose it such that it contains required products, $\mathcal{A}^{(2)}  \subset \mathcal{B}$, being also padded to have a power-of-two cardinality. Next, we demand that there exists a state $|xx\rangle$ that provides such basis, and that the product can be defined via the overlap with some state $| gh \rrangle$,
\begin{align}
    g(x)h(x) &= \langle xx | gh \rrangle = \sum_j (gh)_j \beta_j(x),
\end{align}
where $(gh)_j$ are amplitudes for this state. We note that the encoding circuit for $|xx\rangle$ can differ from $|x\rangle$, however it is often heavily related to $\hat{\mathcal{U}}(x)$ (see the Model Encoding section). Furthermore when $|xx\rangle$ does differ from $|x\rangle$ we still need to be able to represent every term of the DE with the same $x$ dependence. To achieve this any single function terms (e.g. $h(x)$ alone) can be taken and multiplied with $g(x)=1$ as a trivial function. All terms are then expressed via $|xx\rangle$ instead of $|x\rangle$.

Knowing that $g(x)h(x)$ can be expressed in terms of $|xx\rangle$ extended variable state, we need to know how to prepare $|gh\rrangle$. Specifically, this assumes that we have access to circuit that create $|g\rrangle$ and $|h\rrangle$. We note that original product bases in $\mathcal{A}^{(2)}$ can be expanded in the extended $\mathcal{B}$ basis as
\begin{align}
    \tau_j(x) \tau_k(x) = \sum_{l=0}^{2^{\tilde{N}}-1} b^{j,k}_l \beta_l(x),
\end{align}
where $b^{j,k}_l$ are expansion coefficients, and $\tilde{N}$ is the smallest integer that accommodates $\mathcal{B}$ with the specified properties. Substituting this into Eq.~\eqref{eq:prod_fn}, we observe that
\begin{align}
    g(x)h(x) = \sum_{j=0}^{2^N-1} \sum_{k=0}^{2^N-1} \sum_{l=0}^{2^{\tilde{N}}-1} g_j h_k  b^{j,k}_l \beta_l(x) = \sum_{j=0}^{2^{\tilde{N}}-1} (gh)_j \beta_j(x),
\end{align}
and the linear relation between products of amplitudes $\{g_j h_k\}_{j,k}$ and the amplitudes $\{ (gh)_j \}_j$ of the product state can be seen. Therefore, a matrix $\check{M}$ can be constructed such that $|gh\rrangle = \check{M} |g\rrangle |h\rrangle$. Implementation of $\check{M}$ depends on the choice of $\hat{\mathcal{U}}(x)$. Later we describe it for the two specific choices of Fourier and Chebyshev encodings.

Finally, we note that similar strategy can be applied to cases where more that two functions shall be multiplied, where essential steps correspond to defining a suitable variable state in the extended basis, and the corresponding multiply operator.


\subsection{Generalizations}

So far we have described a simple scenario of solving a first order, one-dimensional differential equation. However, many differential equations of interest have more general forms. Next, we briefly discuss how to expand our approach for a wider range of differential equations.

\textit{Introducing nonlinearities.---}In the preceding subsection we have shown that latent space embedded functions can be multiplied in the extended basis. Similar procedure applies for higher-order terms with powers of $f^M(x)$ arranged for extended registers. For equations that include nonlinear function of the model, e.g. $\cos[f(x)]$, we need to use series expansions (Taylor, Fourier, Chebyshev etc), truncating the degree of nonlinearity at some acceptable level.

\textit{Evaluating higher-order derivatives.---}Previously we have suggested that derivatives for overlap models with specific encodings $\hat{\mathcal{U}}(x)$ can be evaluated as $f'_\theta(x) = \langle x | \check{G}^\dag |f_\theta \rrangle$. This simply follows from the fact that $\check{G}$ is a generator for our encoding. Higher-order derivatives follow the same strategy, where we repeatedly ``lower-down'' and concatenate generators, leading to the $m^{th}$-order derivative
\begin{align}
    f^{(m)}_\theta(x) = \langle x | (\check{G}^\dag)^m |f_\theta \rrangle = \langle x | f^{(m)}_\theta \rrangle.
\end{align}
%


\textit{Solving system of differential equations.---}For systems of differential equations, we can treat each separate equation as a part of differential constraints. This leads to the total loss that grows with the number of differential equations, being the sum of contributions,
\begin{align}
    \mathcal{L}_{\mathrm{system}}(\theta) = \sum_{j=1}^D\sum_{k=1}^{T} \llangle \mathrm{DE}_j^{[k]}| \sum_{l=1}^{T} | \mathrm{DE}_j^{[l]} \rrangle.
\end{align}
%


\textit{Tackling multidimensional problems.---}For differential equations that feature several independent variables, generally we need to change the encoding strategy accommodating a vector $\bm{x}$. This can be done by developing encoding circuits for each component, and using the tensor product structure for dealing with them. For instance, consider the simplest generalization to two variables $x$ and $y$. The two-variable model is then instantiated as
\begin{align}
    f_\theta(x, y) = (\langle x|\otimes \langle y|) | f_\theta\rrangle,
\end{align}
and we can extend the tensor product to multiple variables $\{x_i\}_{i}$ with separate registers. When working with these variables we need to employ parallel feature maps, $\bigotimes_i \hat{\mathcal{U}}^i(x_i)$. Correspondingly, the multidimensional transform operator reads $\bigotimes_i \hat{\mathcal{U}}_\mathrm{T}^i$, where $\hat{\mathcal{U}}_\mathrm{T}^i$ is the transform for the $i^{th}$ encoding (which can be different). Next, to take the derivative with respect to the $i^{th}$ variable one needs to act on the $i^{th}$ register with $\check{G}^\dag_i$, being the associated derivative operator for the $i^{th}$ encoding. The multiplier circuit is also constructed depending on the choice of encoding. We also note that one can employ an alternative strategy of encoding multiple variables on the same register using serial circuits (with controls), and leave this as a question for future investigations.


\section{Model Encoding}
Until this point, we have kept the approach general and valid for any choice of encoding that satisfies requirements for the parallel (or global) evaluation. In this section, we introduce two specific example choices of encoding models for global physics-informed quantum DE solvers. For each choice, we detail the components and circuits required for the full workflow. These include feature map $\hat{\mathcal{U}}(x)$, differentiation rules for $d\hat{\mathcal{U}}(x)/dx$ that involve the generator $\check{G}$, transformation circuit $\hat{\mathcal{U}}_\mathrm{T}$, and the multiplication operator $\check{M}$. The choices are the Chebyshev encoding \cite{williams2023quantum} and the Fourier encoding \cite{kyriienko2022protocols}, but we stress that other encodings can be developed in a similar way, for instance the encoding used in recently proposed harmonic neural networks \cite{ghosh2022harmonic}.


\subsection{Chebyshev encoding}

\textit{Feature map.---}The Chebyshev polynomials represent a high-performing basis for regression, function fitting, and integral evaluation tasks \cite{trefethen2019approximation}. They are widely used in spectral methods for solving differential equations, financial modelling, and describing many-body systems. In quantum SciML Chebyshev polynomials were used for solving differential equations with derivative quantum circuits \cite{kyriienko2021solving,Paine2021}, though without linear basis independence. Recently, we have proposed a protocol for generating an orthonormal Chebyshev basis set via quantum feature maps \cite{williams2023quantum}. Building up on the developed toolbox, we show that this basis type enables building global Chebyshev models in the quantum latent space.

We choose the encoding in such a way that amplitudes of a quantum state correspond to $x$-dependent Chebyshev polynomials of the first kind $T_k(x) \equiv \cos(k\arccos{x})$, where $k$ denotes a polynomial degree. The corresponding variable encoding state reads \cite{williams2023quantum}
\begin{align}
|x\rangle = \frac{1}{\mathcal{N}_N(x)}\left(\frac{1}{2^{N/2}}T_0(x)|\mathrm{\o}\rangle + \frac{1}{2^{(N-1)/2}} \sum_{k=1}^{2^N-1} T_k(x)|k\rangle\right),
\end{align}
where we ensure that the states $\{|x\rangle\}$ are orthogonal on the Chebyshev grid with $2^N$ points. This $N$-qubit state is normalized by
\begin{align}
\mathcal{N}_N(x) = \frac{1}{2^{(N-1)/2}}\left(1/2 + \sum_{k=1}^{2^N-1} T_k^2(x) \right)^{1/2},
\end{align}
where we note that in general it is $x$-dependent when evaluated outside of the Chebyshev grid. We want the Chebyshev polynomials themselves to be the basis functions, not rescaled by $\mathcal{N}_N(x)$. Thus, we introduce the corresponding model written in the form
\begin{align}
    f_\theta(x) = \mathcal{N}_N(x) \langle x| f_\theta\rrangle.
\end{align}
The multiplier $\mathcal{N}_N(x)$ leads to a slightly modified workflow concerning post-processing, where relevant overlaps are effectively evaluated with the state $\mathcal{N}_N(x) \langle x|$. However, the encoding of variable $x$ is still based on independent basis functions, and all steps remain valid.


\textit{Derivatives.---}To encode derivatives for Chebyshev-based models, we utilize the beautiful properties of Chebyshev polynomials that involve nesting and shifting rules. Specifically, it is known that derivatives can be expressed as $T_n'(x) = n U_{n-1}(x)$, where $U_n(x)$ are Chebyshev polynomials of the second kind \cite{wang2015some}. This can then be further expanded to
\begin{align}
\label{eq:Cheb-diff-rules-1}
    T_0'(x) &= 0, \\
    T_{2n}'(x) &= 4n \sum_{m=1}^n T_{2m-1}(x), \\
    T_{2n+1}'(x) &= (4n+2) \sum_{m=1}^n T_{2m}(x) + (2n+1)T_0(x). 
\label{eq:Cheb-diff-rules-3}
\end{align}
Therefore, for all Chebyshev polynomials we can write the derivative in an analytical form as a sum
\begin{align}
    T_n'(x) = \sum_{j=0}^{2^N-1} w_j^n T_j(x),
\end{align}
where coefficients $w_j^n$ for each degree are simply collecting the terms from differentiation rules in Eqs.~\eqref{eq:Cheb-diff-rules-1}--\eqref{eq:Cheb-diff-rules-3}.

We reiterate that our goal is creating a derivative state, where amplitudes are now differentiated with respect to $x$. Note that derivatives are exact (no finite differencing is involved), and the procedure can be considered as an automatic quantum differentiation. From this, a derivative operator $\check{G}$ is composed such that
\begin{align}
\frac{d}{dx}\mathcal{N}_N(x)|x\rangle = \check{G} \mathcal{N}_N(x)|x\rangle,
\end{align}
with matrix elements of $\check{G}$ corresponding to
$G_{i,j} = w_j^i c_j$, where $c_0 = \sqrt{2}$,  and $c_j = 1$ for $j \neq 0$. The additional scaling introduced as coefficients $\{c_j\}$ comes from the distinct prefactor of zero degree polynomial $T_0(x)$ (i.e. constant term) compared to other amplitudes of $|x \rangle$, which originates from the Chebyshev orthonormality conditions \cite{williams2023quantum}. Since $\check{G}$ is generally a non-unitary matrix, it can be implemented with the LCU approach, including its ancilla-free versions without increasing the circuit depth, and encoding the action of $\check{G}$ with the help of an ancilla qubit \cite{childs2012hamiltonian, low2019hamiltonian}. 
With this we get the derivative expressed as 
\begin{align}
    f_\theta'(x) = \mathcal{N}_N(x) \langle x| \check{G}^\dag |f_\theta \rangle.
\end{align}


\textit{Transform.---}At the Chebyshev nodes $\{x_j\}_j, ~ x_j = \cos[(2j+1)\pi/2^{N+1}]$, the states prepared by the Chebyshev feature map $\{\mathcal{N}(x_j)|x_j\rangle\}_j$ form an orthonormal basis which we call the Chebyshev basis. Therefore, there exists an operator $\hat{\mathcal{U}}_{\mathrm{QChT}}$ which transforms between the computational basis (real-space grid) and the Chebyshev basis. The corresponding transformation operator is introduced in Ref.~\cite{williams2023quantum} and allows relating functions values at $\{x_j\}_j$ with amplitudes of a state via
\begin{align}
    g(x) = \mathcal{N}_N(x) \langle x | \hat{\mathcal{U}}_{\mathrm{QChT}} \hat{\mathcal{U}}_{\mathrm{QChT}}^\dag | g \rangle.
\end{align}

\textit{Multiplier.---}Next, we discuss how to build the multiplier in the Chebyshev basis and introducing nonlinear terms in differential equations, which is a non-trivial problem. We begin by recalling the multiplication rules of the Chebyshev polynomials,
\begin{align}
    \label{eq:cheb_mult_rules}
    T_j(x) T_k(x) = \Big[T_{j+k}(x) + T_{|j-k|}(x) \Big]/2,
\end{align}
and aim to exploit them for the automatic multiplication. From the relation above we observe that the basis of the product of two functions based on Chebyshev polynomials up to $T_{2^N-1}$ can be expressed as the Chebyshev polynomial up to $T_{2^{N+1}-1}$ degree. Specifically, we can write the product basis as
\begin{align}
    \mathcal{N}_{N+1}(x) | xx\rangle = \frac{1}{2^{(N+1)/2}}T_0(x)|\mathrm{\o}\rangle + \frac{1}{2^{N/2}} \sum_{k=1}^{2^{N+1}-1} T_k(x)|k\rangle,
\end{align}
which is simply the Chebyshev encoding for $N+1$ qubits. Using Eq.~\eqref{eq:cheb_mult_rules}, the desired product state $|gh\rangle$ has to be of the form
\begin{align}
    |gh\rangle_l &= \frac{1}{2^{N/2}}\sum_{j= \max(0, l-2^N+1)}^{\min(l, 2^N-1)} \frac{c_l}{c_{j}c_{l-j}}|g\rangle_{j} |h\rangle_{l-j} \\
    &+ \frac{1}{2^{N/2}} \sum_{j= l}^{2^N-1} \frac{1}{c_{j}c_{j-l}c_l} \left(|g\rangle_{j} |h\rangle_{j-l} + |g\rangle_{j-l} |h\rangle_{j} \right),
\end{align}
where $\bm{c}$ has values $c_0 =\sqrt{2}$ for $c_j = 1~j\neq0$. $\bm{c}$ accounts for the different coefficient of $T_0(x)$.  We need to implement an operator $\check{M}$ that acts as
\begin{align}
    \label{eq:cheb_mult_act}
    \check{M}:  |g \rangle |h\rangle |0 \rangle &\xrightarrow{} \frac{1}{2^{N/2}} |a\rangle \sum_{j,k} \frac{c_{j+k}}{c_j c_k} g_j h_k|j+k \rangle \\
    &+ \frac{1}{2^{N/2}} |\tilde{a}\rangle \sum_{j,k} \frac{c_{|j-k|}}{c_j c_k} g_j h_k||j-k| \rangle,
\end{align}
where $|a\rangle$ and $|\tilde{a}\rangle$ are (combined registers of) ancillary states that can be discarded later.
\begin{figure}
    \begin{center}
    \includegraphics[width=1.0\linewidth]{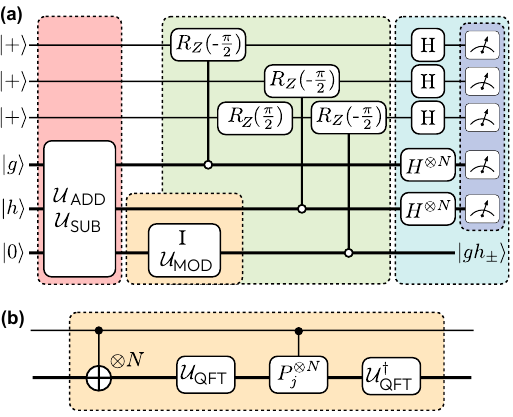}
    \end{center}
    \caption{\textbf{(a)} Circuit diagram of multiplication circuits $\check{M}_+$ and $\check{M}_-$, which together implement the multiplier operator $\check{M}$ for the Chebyshev encoding. First, either $\hat{\mathcal{U}}_{\mathrm{ADD}}$ or $\hat{\mathcal{U}}_{\mathrm{SUB}}$ are applied followed by identity or  $\hat{\mathcal{U}}_{\mathrm{MOD}}$, respectively. Details of $\hat{\mathcal{U}}_{\mathrm{MOD}}$ are shown in figure panel (b). Then, the block of rotations $\hat{C}$ is applied to alter the coefficients. Finally, Hadamards on all ancillary qubits are applied before a projective measurement onto $|0\rangle$. The probability of success of the projective measurement is used to estimate the renormalization factors $\tilde{r}_\pm$. \textbf{(b)} Circuit diagram of $\hat{\mathcal{U}}_{\mathrm{MOD}}$. First, a layer of CNOTs are applied controlled by the most significant bit (MSB). Then via the quantum Fourier transform (QFT) implementation of the adder circuit the MSB is added to the remainder of the register. The phase gate $\hat{P}_j$ corresponds to $\mathrm{diag}([1, \exp(2\pi i/2^j)$, where $j$ is a qubit index.}
    \label{fig:cheb_mult}
\end{figure}
This operator can be split into two parts, where $\check{M}_+$ prepares the summation $|j+k\rangle$ with corresponding amplitude multiplication, and $\check{M}_-$ prepares the summation of $||j-k|\rangle$. For $\check{M}_+$ we use the adder circuit. The adder circuits for  $\check{M}_{\pm}$ can be implemented in several ways \cite{orts2020review, ruiz2017quantum}. We consider the realization, which acts as \cite{wang2016improved, daei2020optimized}
\begin{align}
    \hat{\mathcal{U}}_{\mathrm{ADD}}: |g\rangle |h\rangle |0\rangle \xrightarrow{}  \sum_{j,k} |j\rangle |k\rangle g_j h_k |j+k \rangle.
\end{align}
This is close to our desired operation, and can be used as a base for $\check{M}_+$ (Fig.~\ref{fig:cheb_mult}a). Next, the effects of varying coefficients $\bm{c}$ need to be included. After the adder is applied, an operator $\hat{C}$ is needed which acts according to the rules:
\begin{enumerate}
    \item If $|j\rangle = |0\rangle$ alter shift by $\frac{1}{\sqrt{2}}$.
    \item If $|k\rangle = |0\rangle$ alter shift by $\frac{1}{\sqrt{2}}$.
    \item If $|j+k\rangle = |0\rangle$ alter shift by $\sqrt{2}$.
\end{enumerate}
To implement this we use controlled $\hat{R}_Z$ rotations acting on ancillas (Fig.~\ref{fig:cheb_mult}a, three qubits at the top). The ancillas are prepared in the uniform state. Then, one controlled rotation is used for each shift: $\hat{R}_Z(-\pi/2)$ controlled by $|j\rangle = |0\rangle$, $\hat{R}_Z(-\pi/2)$ controlled by $|k\rangle = |0\rangle$, and $\hat{R}_Z(-\pi/2)$ controlled by $|j+k\rangle = |0\rangle$. The last of which if followed by a single-qubit rotation $\hat{R}_Z(\pi/2)$ (not controlled). Hadamards are then applied to the ancilla and a projective measurement onto $|0\rangle$ reduces the operation to
\begin{align}
    \hat{C}: |0\rangle \sum_{j,k} |j\rangle& |k\rangle g_j h_k |j+k \rangle \xrightarrow{} \\
    &\frac{1}{r_+} |0\rangle \sum_{j,k} |j\rangle |k\rangle \frac{c_{j+k}}{c_j c_k} g_j h_k |j+k \rangle,
\label{eq:entangled}
\end{align}
where $r_+$ is the renormalization factor after the measurement and is accounted for later.

In Eq.~\eqref{eq:entangled} registers $|j\rangle$ and $|k\rangle$ are entangled with $|j+k\rangle$. To discard the loading registers safely, we want to remove this entanglement, and apply a layer of Hadamards to all qubits within $|j\rangle$ and $|k\rangle$ followed by projective measurements onto $|0\rangle$. The corresponding operator $\check{D}$ acts as
\begin{align}
    \check{D}: \frac{1}{r_+}|0\rangle \sum_{j,k} |j\rangle &|k\rangle \frac{c_{j+k}}{c_j c_k} g_j h_k |j+k \rangle \xrightarrow{} \\
    &\frac{1}{\tilde{r}_+} |0\rangle |0\rangle  |0\rangle \sum_{j,k} \frac{c_{j+k}}{c_j c_k} g_j h_k |j+k \rangle,
\end{align}
which is our desired state other than the renormalization factor $\tilde{r}_+$. This constant $\tilde{r}_+$ can be estimated as the square root of the success probability of the projective measurements onto $|0\rangle$. We can now compile $\check{M}_+$ as
\begin{align}
    \check{M}_+ = \tilde{r}_+ \check{D} \hat{C} \hat{\mathcal{U}}_{\mathrm{ADD}} .
\end{align}

The subtraction part $\check{M}_-$ of the multiplier is prepared in similar way. We start with a subtractor circuit as a base,
\begin{align}
    \hat{\mathcal{U}}_{\mathrm{SUB}}: |g\rangle |h\rangle |0\rangle \xrightarrow{}  \sum_{j,k} |j\rangle |k\rangle g_j h_k |j-k ~ \mathrm{mod} ~2^{N+1} \rangle.
\end{align}
A subtractor circuit is generally an adapted adder circuit, as subtraction is the inverse operation of addition. However, an extra operator $\hat{\mathcal{U}}_{\mathrm{MOD}}$ is now needed (Fig.~\ref{fig:cheb_mult}b), as we want to prepare $||j-k|\rangle$ but the subtractor prepares $|(j-k)~\mathrm{mod} ~2^{N+1} \rangle$. Therefore, when $(j-k)<0$ we need to map states from $|(j-k)~ \mathrm{mod} ~2^{N+1} \rangle = |2^{N+1}+j-k\rangle$ to $|k-j\rangle$. Since $|j-k|<2^N$, when using $N+1$ qubits we can separate the cases of $(j-k)<0$ and $(j-k) \geq 0$ by the value of the most significant bit (MSB) in the result register. Namely, we get $|1\rangle$ in MSB if $(j-k)<0$, and $|0\rangle$ otherwise. To implement the mapping, a layer of CNOTs controlled by the MSB is applied to the result register, performing $|2^{N+1}+j-k\rangle \rightarrow |k-j-1\rangle $. Then, controlled again by the MSB in $|1\rangle$, the state is added with the adder circuit to the result register for $||j-k|\rangle$,
\begin{align}
    \hat{\mathcal{U}}_{\mathrm{MOD}}: \sum_{j,k} |j\rangle &|k\rangle g_j h_k |j-k ~ \mathrm{mod} ~2^{N+1} \rangle \xrightarrow{} \\
    & \sum_{j,k} |j\rangle |k\rangle g_j h_k | |j-k|  \rangle.
\end{align}
For the rest of the circuit, we also need to introduce rescaling to account for the differing coefficient of $T_0$. The operator $\hat{C}$ can be used same way as before. Also similar to the previous case, the other registers need to be disentangled from the result register with Hadamards and projective measurements, with the renormalization factor $\tilde{r}_-$. This operator $\check{\tilde{D}}$ slightly differs from $\check{D}$ in that it must also apply to the MSB of the results register. This introduces an extra scale factor of $2$. Summarizing these steps, we have
\begin{align}
    \check{\tilde{D}}\hat{C}: &\sum_{j,k} |j\rangle |k\rangle g_j h_k | |j-k|  \rangle \xrightarrow{} \\
    &\rightarrow \frac{2}{\tilde{r}_-} |0\rangle |0\rangle \sum_{j,k} \frac{c_{|j-k|}}{c_j c_k} g_j h_k ||j-k| \rangle, \\
    \check{M}_- &= \frac{\tilde{r}_-}{2} \check{\tilde{D}} \hat{C} \hat{\mathcal{U}}_{\mathrm{MOD}} \hat{\mathcal{U}}_{\mathrm{SUB}}.
\end{align}
The entire multiplier operator can be pulled together as
\begin{align}
    \check{M} = \frac{1}{2^{N/2}} \left(\check{M}_+ + \check{M}_- \right),
\end{align}
and is shown in Fig.~\ref{fig:cheb_mult}.


\subsection{Fourier encoding}

\textit{Feature map.---}Another popular basis for building various mathematical models is the Fourier basis. This basis is formed of the functions $T^N_j(x) \coloneqq \exp(i 2 \pi j x/ 2^N )$, where $N$ is the number of qubits. The corresponding feature map with exponentially large capacity is referred to as the phase feature map, introduced in Ref.~\cite{kyriienko2021solving}. The phase feature map is formed by a layer of Hadamards on each qubit followed by a layer of controlled phase gates, $\hat{P}_N(x, j) = \mathrm{diag}\{1, \exp(i \pi x 2^{j-N})\}$, where $j$ is a qubit index we act upon. Applied to the computational zero, the phase feature map prepares the state
\begin{align}
    |x\rangle = \frac{1}{2^{N/2}}\sum_{j=0}^{2^N-1}T^N_j(x) |j\rangle.
\end{align}
The set of corresponding Fourier states $\{|x_j\rangle \}_j$ evaluated at integer points $\{j\}_{j=0}^{2^N-1}$ is orthonormal.

\textit{Derivatives.---}When taking derivatives we observe 
 that
\begin{align}
   \hat{P}_N'(x,j)  =& ~\mathrm{diag}(0, i \pi 2^{j-N})P(x, j) = \check{G}_j \hat{P}_N(x, j), \\
   \check{G}_j =& i \pi 2^{j-N-1} \hat{I} - i \pi 2^{j-N-1} \hat{Z}.
\end{align}
Using the product rule, we repeat the procedure for other gates, leading to $\check{G} = \sum_j \check{G}_j$ as the full derivative operator for the Fourier encoding.

\textit{Transform.---}We have already recalled that at the set of points $\{j\}_{j=0}^{2^N-1}$ the Fourier map prepares an orthonormal basis. This basis can be mapped to the computational basis by a unitary operator (bijection). The corresponding transform operator is the quantum Fourier transform (QFT) \cite{nielsen2010quantum}. 


\textit{Multiplier.---}When considering a multiplication operator, we again begin by recalling the multiplication rules of the basis functions. In the Fourier case these simply correspond to
\begin{align}
\label{eq:fourier_mult_rule}
    T^N_j(x) T^N_k(x) &= T^N_{j+k}(x).
\end{align}
The resulting product basis state reads
\begin{align}
|xx\rangle &= \frac{1}{2^{(N+1)/2}}\sum_{j=0}^{2^{N+1}-1} T_j^N(x).
\end{align}
The form of $|xx\rangle$ is similar to $|x\rangle$, and the same encoding circuit can be used when extended to $N+1$ qubits by acting on the additional qubit with $P_N(x, N+1)$. From this we find the desired $|gh\rangle$ and the multiplication operator $\check{M}$:
\begin{align}
    &|gh\rangle_l = \frac{1}{2^{(N-1)/2}} \sum_{j= \max(0, l-2^N+1)}^{\min(l, 2^N-1)} |g\rangle_j |h\rangle_{l-j}, \\
    &\check{M}: |g\rangle |h\rangle |0\rangle \xrightarrow{} \frac{1}{2^{(N-1)/2}} |a\rangle \sum_{j,k} g_j h_k |j+k \rangle.
\end{align}
Comparing to the Chebyshev encoding case in Eq.~\eqref{eq:cheb_mult_act}, we see that the Fourier multiplicator is a simplified version of the Chebyshev multiplicator, as it does not require rescaling and subtraction. The Fourier multiplicator reads
\begin{align}
    \check{M} = \frac{1}{2^{(N-1)/2}} \tilde{r}_+ \check{D} \hat{\mathcal{U}}_{\mathrm{ADD}}.
\end{align}
With this, the necessary building blocks for implementing training with Fourier encoding have been introduced.


\subsection{Comparing the encodings}

Comparing Chebyshev encoding versus Fourier encoding, we note their most notable differences. The first is that the Chebyshev encoding state is real for all $x$ (i.e. it leads to states with real amplitudes). Therefore a purely real function can easily be enforced by using a variational ansatz which prepares states with real-only amplitudes (for instance, based on $\hat{R}_Y$ rotations and fixed CZ/CNOT gates). As many problems considered are purely real, this can be a useful restriction to be able to enforce. The Fourier encoding state is naturally complex, and therefore such a simple enforcement for purely real functions is not available. Potentially, a specific variational ansatz could be created for this purpose but it is not currently known.
\begin{figure}
    \begin{center}
    \includegraphics[width=1.0\linewidth]{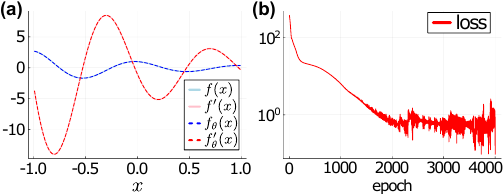}
    \end{center}
    \caption{Results of solving a linear differential equation \eqref{eq:DE-example-1}, for $\kappa = 1$, $\lambda = 2\pi$, and initial value of $f(0) = 1$. \textbf{(a)} Target function and its derivative (solid curves) plotted as a function of $x$, vs parallelly-learnt function and its derivative evaluated from quantum circuits (dash curves). The functions overlay. \textbf{(b)} Value of loss function over epoch (iteration of the Adam optimizer).}
    \label{fig:res_lin}
\end{figure}

The next point of comparison is the basis functions themselves. Fourier encoding basis functions are periodic in nature and are particularly suited to harmonic problems. Chebyshev encoding basis functions are real polynomials, and are known to offer best-in-class series expansion in terms of $L_{\infty}$ norm. For each problem an appropriate basis (one of these two or another entirely) would have to be found. Finally, we note that in general the required operator toolbox for implementing the Fourier encoding is slightly simpler than for the Chebyshev encoding.

With two specific encodings considered, we now reconsider the general case. For any encoding chosen the appropriate operator toolbox must be found. Generally, their implementation will vary greatly based on the encoding. The multiplier in particular will vary. However, we believe that many encodings of interest will have multipliers that are able to be implemented with modifications of the adder and subtractor circuits.


\section{Results}

Now, let us apply the developed strategy to several exemplary differential equations. Here we aim to test the main aspects of the protocol, and do not concentrate on the particular industrial use-cases. These will be considered in a separate publication.


\subsection{Linear differential equation example} 

First, we consider a linear differential equation of the form
\begin{align}
\label{eq:DE-example-1}
    \frac{df(x)}{dx} + \kappa \exp(-\kappa x) \cos(\lambda x) + \lambda \exp(-\kappa x) \sin(\lambda x) = 0,
\end{align}
for some real-valued parameters $\lambda$, $\kappa$, and the initial value $f(0) = 1$. This differential equation has a known analytic solution $f(x) = \exp(-\kappa x)\cos(\lambda x)$, which represents dynamics of a damped oscillator.

We simulate the proposed algorithm, implementing the steps discussed in the previous sections. We use the \emph{Julia} programming language and the \emph{Yao.jl} package \cite{luo2020yao} for the full statevector simulation. In particular we use the Chebyshev encoding and represent the function using four qubits. The derivative operator, multiplier and transform follow the Model Embedding section. For building the model we prepare $|f_\theta\rrangle = \theta_s \hat{\mathcal{U}}_\theta|{\o}\rangle$, where $\theta_s$ is a classical variational parameter representing a scale factor, and $\mathcal{U}_\theta$ is an adjustable circuit (variational ansatz). For the variational ansatz we use alternating layers of $\hat{R}_Y$ rotations and layers of CNOTs. Specifically, seven rotational and six entanglement layers are used, initialized randomly. This ansatz prepares states with real-only amplitudes, leading to functions that are guaranteed real-valued. 
\begin{figure}[t!]
    \begin{center}
    \includegraphics[width=1.0\linewidth]{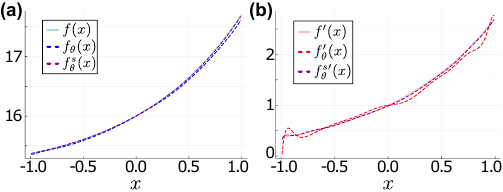}
    \end{center}
    \caption{Results of solving linear differential equation $df(x)/dx - f(x) + 15 = 0$ with initial value of $f(0) = 16$, approached with two different models [Eqs.~\eqref{eq:f-model_sc} and \eqref{eq:f-model_sh}]. \textbf{(a)} Plot of the ground truth function (solid curve) vs trained solution regular [$f_{\theta^*}(x)$] and shifted [$f_{\theta^*}^{\mathrm{sh}}(x)$] models shown by dash curves. \textbf{(b)} Derivative of the ground truth (solid curve) vs derivatives of solutions obtained from training the regular and shifted models (dashed curves). The shifted model closely follows the ground truth, while the regular scaled model starts to deviate from the ground truth, which is evident from derivatives.}
    \label{fig:res_lin_comp}
\end{figure}

The differential equation in the latent space is formed of two terms: the function derivative $df/dx$ and the function of an independent variable $g(x) \coloneqq \exp(-\kappa x) \left(\kappa \cos(\lambda x) + \lambda \sin(\lambda x)\right)$. The latent representation of $df/dx$ is $ |dfdx \rrangle \coloneqq \theta_s \check{G}^\dag \hat{\mathcal{U}}_\theta|0\rangle$. The representation of $g(x)$ in latent space is obtained from the use of the transform, as described in the previous section. We assume the non-normalized version of this wavefunction can be written as a normalized part $|g\rangle$ multiplied with some scalar $N_g$, to attain the right overall scaling in the latent space representation. Aiming to equate the derivative and the independent function, the differential equation loss reads
\begin{align}
    \mathcal{L}_{\mathrm{DE}}(\theta) = &\theta_s^2\langle 0| \hat{\mathcal{U}}^\dagger_\theta \check{G} \check{G}^\dagger \hat{\mathcal{U}}_\theta|0\rangle + \theta_s N_g \langle g|\check{G}^\dagger \hat{\mathcal{U}}_\theta|0\rangle \\
    &+ \theta_s N_g \langle 0| \hat{\mathcal{U}}^\dagger_\theta \check{G} |g \rangle + N_g^2 \langle g| g \rangle,\nonumber
\end{align}
and the initial value component of the loss is $\mathcal{L}_{\mathrm{init}}(\theta) = (f_\theta(0)-1)^2$. 
The total loss is then taken as $\mathcal{L}(\theta) = (\mathcal{L}_{\mathrm{DE}})^p + \eta \mathcal{L}_{\mathrm{init}}$, where we set the weight factor for the initial value loss $\eta=10$ to regularize solutions and emphasize the importance of the boundary. This hyperparameter can be chosen heuristically or meta-trained, and furthermore it can be a function of the epoch number \cite{kyriienko2021solving,Paine2021}. Additionally, we have introduced $p$ as a power for scaling the DE loss, which is chosen as $p=1/2$. The square root is monotonically increasing over $\mathbb{R}_+$, and does not change the optimum. However, it allows converting the loss from MSE to RMSE (root-mean-square error) loss. This is commonly used in deep learning, being more sensitive to outliers (or initial value pinning as in our case). The loss $\mathcal{L}(\theta)$ is minimized via the Adam optimizer with a small learning rate $0.005$ and increasing number of epochs. The results are shown in Fig.~\ref{fig:res_lin}. In Fig.~\ref{fig:res_lin}a the dashed curves show functions and derivatives obtained from $f_{\theta^*}(x)$ and $df_{\theta^*}(x)/dx$, evaluated at quasi-optimal angles $\theta^*$ coming from the optimization procedure. The corresponding decrease of loss as a function of epoch is shown in Fig.~\ref{fig:res_lin}b. We observe that the obtained solutions follow the ground truth for the system (Fig.~\ref{fig:res_lin}a, solid curves).

\subsection{Shifted linear differential equation example}

Next, we address an example that requires extending the expressivity of the model, also improving the trainability. Expressivity and trainability of this model mainly depend on choices of the encoding for $|x\rangle$. However, this can be aided by shifting and scaling, such that we use trainable quantum states to reproduce only a nontrivial $x$ dependence. In the Protocol section we introduced the modified models \eqref{eq:f-model_sc} and \eqref{eq:f-model_sh}. Here, we formulate the model $f^{\mathrm{sh}}_\theta(x) = \theta_s \langle x | \hat{\mathcal{U}}_\theta|{\o}\rangle + \theta_{\mathrm{sh}}$ with variationally adjustable scaling and shifting parameters ($\theta_s$ and $\theta_{\mathrm{sh}}$). This model performs well when the solution value range is significantly smaller than the magnitude of the solution (effectively decreasing the demands for expressivity and boosting trainability). 

To showcase the use of shifted models, we simulate solving the differential equation of the form
\begin{align}
    \frac{df(x)}{dx} - f(x) +15 =0, \quad f(0) = 16,
\end{align}
with both types of models \eqref{eq:f-model_sc} and \eqref{eq:f-model_sh}. This problem has an analytic solution $f(x) = \exp(x)+15$ as a ground truth. We use Chebyshev encoding over four qubits and the real-only ansatz of depth six as used in previous results. With the Chebyshev encoding $g(x) = 1$ is an accessible basis function, therefore the appropriate state representation $\psi_1$ can be found by inspecting basis functions or using the transform.

For this differentiation equation there are three DE terms: $df/dx$ is represented by $|\mathrm{DE}_1\rangle = \theta_s \check{G}^\dag \hat{\mathcal{U}}_\theta |{\o}\rangle$; $f$ is represented by $|\mathrm{DE}_2\rangle = \theta_s \hat{\mathcal{U}}_\theta |{\o}\rangle$ for the regular scaled model in Eq.~\eqref{eq:f-model_sc}; and $|\mathrm{DE}_2\rangle = \theta_s \hat{\mathcal{U}}_\theta |0\rangle + \theta_{\mathrm{sh}} |\psi_1\rangle$ for the shifted model in Eq.~\eqref{eq:f-model_sh}. The constant function $g(x)=15$ is represented by $|\mathrm{DE}_3\rangle = 15 |\psi_1 \rangle$. The differential loss is them formed as in Eq.~\eqref{eq:loss_DE}. We also include the initial value loss as $\mathcal{L}_{\mathrm{init}}(\theta) = \{f_\theta(0)-16\}^2$.
\begin{figure}
    \begin{center}
    \includegraphics[width=1.0\linewidth]{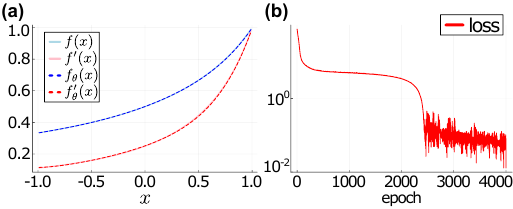}
    \end{center}
    \caption{Results of solving nonlinear DE $df/dx - f^2 = 0$ with initial value $f(0) = 0.5$. \textbf{(a)} Plot of target function and derivative (solid lines) versus resulting function and derivative (dash lines). \textbf{(b)} Value of loss function over epoch iteration.}
    \label{fig:res_nonlin}
\end{figure}

We follow the same training workflow as in the previously considered linear case. We rescale and combine loss contributions as $\mathcal{L}(\theta) = (\mathcal{L}_{\mathrm{DE}})^p + \eta \mathcal{L}_{\mathrm{init}}$, with $p=1/2$ and $\eta=10$. The loss is minimized with Adam optimizer with the learning rate of $0.005$ for both models. Results are shown in Fig.~\ref{fig:res_lin_comp}. We observe that the shifted model performs better (purple dashed curves in Fig.~\ref{fig:res_lin_comp}a,b for function and derivative). We achieve a better performance with shifting, and require fewer epochs roughly reduced by an order of magnitude, signifying the improved trainability.


\subsection{Nonlinear differential equation example} 
We now move on to consider an example of a nonlinear differential equation, chosen as
\begin{align}
\label{eq:DE-nonlinear}
    \frac{df(x)}{dx} + f^2(x)= 0, \quad f(0) = 1/2.
\end{align}
This DE has an analytic solution $f(x) = 1/(2-x)$, set as the ground truth. Notably, even though Eq.~\eqref{eq:DE-nonlinear} looks simple, we should appreciate that solving nonlinear differential equations with amplitude encoding on quantum computing is far away from being easy. In our approach we are able to tackle it with latent space basis multipliers.

We solve the problem following the workflow used in the linear case, while changing the variable state representation and using multiplier circuits $\check{M}$. We employ the Chebyshev encoding on three qubits and the same depth-six variational ansatz. The two terms of this DE, $df/dx$ and $f^2$, are then mapped into the latent space with $|xx\rangle$ variable encoding. The first term is prepared by multiplying $g(x) = 1$ and $df/dx$ for $|\mathrm{DE}_1\rrangle = \theta_s \check{M}(|g\rangle \otimes \check{G}^\dag \hat{\mathcal{U}}_\theta |{\o}\rangle)$. As unity is a naturally occurring basis function of the Chebyshev encoding, the relevant $|g\rangle$ can easily be found as $2^{N/2}[1, 0, 0, ...]$. The second term $f^2$ is represented by $|\mathrm{DE}_2\rrangle = \theta_s^2 \check{M} (\hat{\mathcal{U}}_\theta |{\o}\rangle \otimes \hat{\mathcal{U}}_\theta |{\o}\rangle)$. These states are used to form a differential equation loss  in Eq.~\eqref{eq:loss_DE}, supplemented by initial value loss as $\mathcal{L}_{\mathrm{init}}(\theta) = \{f_\theta(0) -  1/2\}^2$. The total loss is the same as for previous problems, $\mathcal{L}(\theta) = (\mathcal{L}_{\mathrm{DE}})^p + \eta \mathcal{L}_{\mathrm{init}}$, $p=1/2$, $\eta=10$. The loss is minimized with Adam optimizer with the learning rate of $0.005$. The resulting solutions are shown in Fig.~\ref{fig:res_nonlin}a, where the trained solution of the nonlinear DE (blue and red dashed curves for the function and its derivative, respectively). This follows nicely the ground truth (solid curves in Fig.~\ref{fig:res_nonlin}a). The corresponding training is presented in Fig.~\ref{fig:res_nonlin}b, where we observe that the optimizer explores the landscape of solutions, and ultimately finds the suitable candidate when loss drops below unity.


\subsection{Multidimensional differential equation example}

Finally, we test the solver for tackling multidimensional problems. We consider a differential equation for a function $f(x,y)$ of two independent variables $x$ and $y$, written as
\begin{align}
    \frac{df(x,y)}{dy} - 2y - x = 0, \quad f(x, 0) = 1.
\end{align}
\begin{figure}
    \begin{center}
    \includegraphics[width=1.0\linewidth]{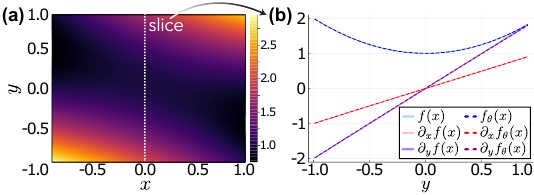}
    \end{center}
    \caption{Results of solving a multidimensional differential equation $df/dy -2y -x = 0$ with the boundary condition $f(x,0) = 1$. \textbf{(a)} Result from parallel multidimensional training shown as a density plot of $f_\theta(x,y)$ vs $x$ and $y$. \textbf{(b)} Slice of resulting function and derivatives at $x=0$. Analytic solutions shown in solid lines. Functions resulting from training shown in dashed lines.}
    \label{fig:res_multidim}
\end{figure}
This DE has an analytic solution $f(x,y) = y^2 + x y + 1$ as the ground truth.

We encode the two independent variables in parallel registers using the model
\begin{align}
    f_\theta(x,y) = \theta_s (\langle x | \otimes \langle y|) |f_\theta \rangle.
\end{align}
We choose to encode both $|x\rangle$ and $|y\rangle$ with Chebyshev encoding of two qubits. $|f_\theta \rangle$ is prepared with the same real-only ansatz as in previous examples, applied over four qubits in total, with an ansatz depth of six. 
The two terms of the differential equation are $df/dy$ and $g(x,y) = -2y -x$. For the derivative term, as variables are encoded in separate registers, the derivative operator $\check{G}^\dag$ is applied to the register of the variable being differentiated. The remaining register is left untouched. Therefore, the corresponding state is $|\mathrm{DE}_1 \rrangle = \left(I \otimes \check{G}^\dag \right) \hat{\mathcal{U}}_\theta |{\o}\rangle$. For the second term, the latent space state is prepared by applying $\hat{\mathcal{U}}^\dag_{\mathrm{QChT}} \otimes \hat{\mathcal{U}}^\dag_{\mathrm{QChT}}$ to the state with amplitudes of $g(x,y)$ evaluated at the Chebyshev nodes for $|\mathrm{DE}_2\rrangle$. Next, we need to sum the differential term and the boundary condition term, $\mathcal{L}_{\mathrm{BC}}(\theta) = \sum_j \{ f_\theta(x_j, 0) - x_j \}^2$, where $\{x_j\}$ are a set of boundary evaluation points. We choose $21$ uniformly spaced points for $x \in (-1,1)$. Again, the total loss is $\mathcal{L}(\theta) = (\mathcal{L}_{\mathrm{DE}})^p + \eta \mathcal{L}_{\mathrm{BC}}$, $p=1/2$, $\eta = 10$. Minimizing this loss via Adam optimizer with the learning rate of $0.005$ we get results plotted in Fig.~\ref{fig:res_multidim}. The full trained solution is shown as a density plot in Fig.~\ref{fig:res_multidim}a, closely following the ground truth (average deviation is in the range of $10^{-3}$). We also take a slice at $x=0$ and plot the solution as a function of $y$ in Fig.~\ref{fig:res_multidim}b. We recover the expected behavior from the multidimensional training.



\section{Discussion}

An important question is about the scaling of the proposed protocol, and its place in the families of quantum and classical differential equation solvers. 

First, we highlight that our approach is developed as a general paradigm for representing physics-informed constraints, and in practice can be implemented with different workflows (involving either variational training, or linear equation solvers based on matrix inversion). Depending on the chosen workflow, one can make it more suitable for noisy intermediate scale quantum (NISQ) devices \cite{Preskill2018quantumcomputingin}, or for \textit{early} fault-tolerant devices, which was recently described as the ISQ era \cite{ISQ} (simply dropping the ``noisy" adjective from NISQ). A significant benefit of our approach corresponds to the global constraint evaluation, where the dependence on the size of the collocation point grid is alleviated. Additionally, the expressivity of the proposed encoding allows compressing solutions to a smaller number of qubits (logarithmic in the number of basis functions needed), and the ability to implement initial value constraints when choosing the variational workflow. This comes at the expense of non-convex optimization (similarly to deep learning approaches), which can be challenging when learning with quantum circuits. We note that training issues such as barren plateaus \cite{mcclean2018barren,Cerezo2021NatComm} and local minima \cite{you2021exponentially,Anschuetz2022} are generally applicable to all protocols that are not tailored to specific problems. However, careful ansatze construction, advanced optimization, and favorable generalization can help to boost the trainability \cite{pesah2021absence,larocca2023theory,Caro2022,holmes2022connecting,Boyd2022CoVaR}. Beyond the state preparation, we also note that multiplication, differentiation and mapping circuits add overheads that may be difficult to handle within NISQ. But given that they are mostly involving QFT-like circuits with $\mathcal{O}(N^2)$ gatecount or even more resource-frugal approximators \cite{Nam2020aqft}, the proposed circuits fit the ISQ requirements. Furthermore, at large scale the inversion-based methods shall become feasible \cite{harrow2009quantum,childs2017quantum,Lin2020optimalpolynomial}, which can work together well with the strategies proposed here, bypassing training at the expense of larger depth, condition number scaling, and difficulty in implementing data-based terms.

Next, we shall try to position the proposed approach among other strategies for solving nonlinear differential equations. In essence, our approach corresponds to a mixture of spectral methods \cite{boyd2013chebyshev,hussaini1983spectral} with the physics-informed neural network structure \cite{Raissi2019PINNs,kyriienko2021solving}. Standard PINNs supply models in terms of nonlinearly-transformed basis functions, which do not need to be linearly independent, and for training they require the full collocation grid evaluation with the size $L_c$ chosen to avoid underfitting and overfitting. Spectral methods however work with orthonormal basis sets (with a prominent example of \emph{chebfun} \cite{trefethen2019approximation}), where the grid is fixed by the degree of polynomials used in representing solutions. In our case, we do require the linear basis function independence that enables the global (overlap-based) comparison of differential equation terms. In this sense, the connection to spectral methods is stronger, and distantly resembles classical Spectral PINNs \cite{lutjens2021spectral} and DeepONets \cite{Meuris2023deeponets}. Also, much like for spectral methods, the models that we build depend on the chosen basis sets. This is particularly important when taking derivatives, as effectively we use continuation outside of the nodes for an underlying computational grid. However, there are also significant distinctions. Given the mapping between spectral model representation and computation basis by unitary transforms, we have fast access to the computational grid evaluation (both explicit and implicit). Next, our approach to multivariate functions and inherently exponential model capacity makes it suitable for problems in higher dimensions. Specifically, for $L_c$ basis functions (and even more collocation points used in PINNs) we need $\mathcal{O}(\log L_c)$ per dimension, and $\mathcal{O}(D \log L_c)$ qubits for $D$-dimensional problems. The differential loss based on overlaps also suits the quantum model discovery framework \cite{heim2021quantum}, while the generator-based differentiation also gives direct access to integral transforms \cite{Kumar2022}. Finally, we can build-in the nonlinearity into our solver, avoiding linearization or convolution that is typically employed in spectral methods. 

Comparing to other \textit{quantum} solvers, we note that ours is distinct from the grid discretization-based approaches \cite{Montanaro2016,Costa2019,Costa2022}. Indeed, even if evaluation is restricted to the computational grid, we go beyond it when dealing with derivatives. Another distinction is the way nonlinearity is introduced into the system \cite{JPLiu2021,Krovi2023improvedquantum}. When comparing to variational approaches with the amplitude encoding \cite{lubasch2020variational,Garcia-Molina2022} we observe that they either use expectation value-based loss or utilize time-stepping, thus being limited to simulating dynamics. Another significant difference with the introduction of nonlinearity via quantum nonlinear processing units that require system size doubling for introducing quadratic nonlinearity (multiplicative cost in the degree of nonlinearity), over the linear system size increase that we require in the latent space. This additive cost depends on the choice of encoding, and can be just a single qubit for the specific encodings introduced. Finally, when compared to the differential quantum circuits approach (DQC) with non-restricted feature maps \cite{kyriienko2021solving,Paine2021,heim2021quantum}, as already mentioned, we do not require grid evaluation in training, thanks to specifically chosen encoding types and overlap evaluations.


\section{Conclusion}

In this work we have developed an approach to solving differential equations by introducing physics-informed constraints into quantum models. The core part of our approach is the latent space representation of terms in differential equations, where functions and derivatives can be effectively represented by quantum states. This is achieved with the use of basis set decomposition with linearly independent basis functions, which depends on a considered feature map. The physics-informed constraints are then implemented as overlaps between quantum states (effective distances between representations of differential equation terms), which are combined into a loss that can be minimized variationally, or a set of equations that can be solved with quantum matrix inversion methods. We have also shown how mapping to explicit models can help encoding initial and boundary terms, as well as adding data-based constraints. Crucially, we present tools for implementing nonlinear terms and multidimensional encoding in the latent space, as well as utilizing automatic model differentiation specific to the chosen basis (in practice these are taken as Chebyshev and Fourier encodings). Simulating the protocols, we solve various exemplary problems that include linear, nonlinear and multidimensional differential equations. Further research could investigate in more detail the potential for quantum computing, using this method, to improve on the limiting factors of PINN methods with regards to massive grid evaluations in higher dimensions.

\begin{acknowledgments}
We thank Chelsea A. Williams, Hsin-Yu Wu, and Atiyo Ghosh for discussions on the subject.
\end{acknowledgments}



\appendix

\section{\label{app:A}Overlap Measurement}
\begin{figure}
    \begin{center}
    \includegraphics[width=1.0\linewidth]{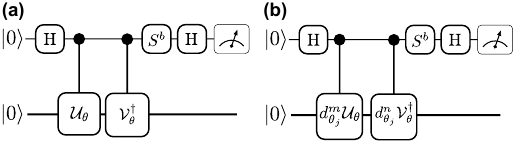}
    \end{center}
    \caption{Circuit diagram of Hadamard tests to measure overlaps and derivatives. \textbf{(a)} Hadamard test for measuring Re$\big\{\langle {\o} |\hat{\mathcal{U}}_\theta^\dag \hat{\mathcal{V}}_\theta | {\o} \rangle \big\}$ and Im$\big\{\langle {\o} |\hat{\mathcal{U}}_\theta^\dag \hat{\mathcal{V}}_\theta | 0{\o} \rangle \big\}$ for $b=0$ and $b=1$, respectively. $\hat{S}$ denotes the phase gate, defined as $\hat{S} = \mathrm{exp}(-i \pi \hat{Z}/4)$. \textbf{(b)} Using Hadamard test for evaluation of the overlap $\frac{d}{d\theta_j}\langle {\o} | \hat{\mathcal{U}}_\theta^\dagger \hat{\mathcal{V}}_\theta | {\o} \rangle$, where $m$ and $n$ indices run over which gates with $\theta_j$ as parameters that are differentiated. When $b=0$ and $b=1$ are used, real and imaginary parts are evaluated, respectively. By summing over $j,k$ and $b = 0,1$ the full overlap can be estimated.}
    \label{fig:overlap_m}
\end{figure}

Here we give a reminder on how to measure overlaps with quantum circuits. As we are not interest in the absolute value squared of state overlaps, but their individual real and imaginary parts, the workflow shall be modified slightly modified from standard test. The relevant circuits are shown in Fig.~\ref{fig:overlap_m} representing the Hadamard tests with a possible addition of the phase gate $\hat{S}$ \cite{Mitarai2019Htest}. First, the ancilla register is put into the symmetric superposition state, followed by the two controlled unitaries (one being reverted). Upon measurement in the Pauli $\hat{X}$ basis (with and without prior phase rotation), we get the real and imaginary parts of overlaps, correspondingly (Fig.~\ref{fig:overlap_m}a). The derivative-based overlaps can be evaluated in the similar fashion (Fig.~\ref{fig:overlap_m}b).

\section{\label{app:B}Alternative variationless workflow}
We consider an alternative to the variational procedure, using the same model specification and understanding of operators representing differential equations terms. Following the steps in Sec. IIB of the main text, the problem can be written as a system of equations corresponding to
\begin{align}
    \sum_{k=1}^T |\mathrm{DE}^k\rrangle = \bm{0}.
\end{align}
First, assuming a linear problem with no cross products between functions of the independent variable and the model (trial function $\bm{f}$), the terms of the DE can be separated into those that depend on the trial function and those that do not. All terms depending on $\bm{f}$ can be represented by a matrix acting on the components of the trial function, $D_j\bm{f}$. The terms that do not depend on $\bm{f}$ can be written directly as a vector of corresponding components, $\bm{g}_j$. Thus we get the equivalent system of equations
\begin{align}
    \sum_j D_j \bm{f} = \sum_j \bm{g}_j \xrightarrow{} D \bm{f} = \bm{g}.
\end{align}
Importantly, we also need to introduce the initial value $f(x_0)=f_0$. For this we observe that $\langle x_0 | \bm{f} = f(x_0) = f_0$. Therefore, to account for an initial value when solving the problem $\langle x_0 |$ is appended as an additional row of $D$ and $f_0$ appended to $\bm{g}$. This matrix equation can then be solved with an appropriate linear systems of equations (LSE) solver. Solving LSEs can be approached with paradigmatic quantum algorithms such as HHL \cite{harrow2009quantum,Biamonte2017} and adiabatic inversion \cite{subacsi2019quantum,Costa2022}, as well as their variations. This is promising in terms of big-O scaling, while may lead to significant circuit depth in practice. The result from HHL and general LSE solvers is the prepared state $|\bm{f}\rangle$ which satisfies the solution for the problem. This can then be measured as an overlap with $|x\rangle$ to be able to measure the resulting function at any point $x$. The full details of such an approach and generalizing is left for future research. In the present paper we focused on learning based on the variational approach, as it maintains the benefits of parallel solving along with the flexibility that differential quantum circuits have due to regularization, data-based constraints, and symmetry embedding.


\input{parallel-DQC.bbl}

\end{document}

%% file: parallel-DQC.bbl
%